\documentclass[preprint2]{aastex}
\shorttitle{GJ867b}
\shortauthors{Davison}
\usepackage{rotating}
\usepackage{pdflscape}
\usepackage{epsfig}
\usepackage{natbib}
\usepackage{subfigure}

\begin{document}

\title{The Closest M-Dwarf Quadruple System to the Sun}
\author{Davison, Cassy L.$^1$, White, R.J.$^1$, Jao, W-C.$^1$, Henry, T.J.$^1$, Bailey, J.I., III$^2$, Quinn, S.N.$^1$, Cantrell, J.R.$^1$, Riedel, A.R.$^{3,4}$, Subasavage, J.~P.$^5$, Winters, J. G.$^1$,  and Crockett, C.J.$^5$}
\affil{$^1$Georgia State University, Department of Physics and Astronomy, Atlanta, GA, 30303, USA}
\affil{$^2$University of Michigan, Department of Astronomy, Ann Arbor, Mi, 48109, USA}
\affil{$^3$Department of Physics and Astronomy, Hunter College, New York, NY, 10065, USA}
\affil{$^4$American Museum of Natural History, New York, NY, 10024, USA}
\affil{$^5$United States Naval Observatory, Flagstaff, AZ, 86002, USA}

\begin{abstract}  
We report new infrared radial velocity measurements obtained with CSHELL at NASA's Infrared Telescope Facility that reveal the M3.5 dwarf GJ 867B to be a single-lined spectroscopic binary with a period of 1.795 $\pm$ 0.017 days.  Its velocity semi-amplitude of 21.4 $\pm$ 0.5 kms$^{-1}$ corresponds to a minimum mass of 61 $\pm$ 7 M$_{JUP}$; the new companion, which we call GJ 867D, could be a brown dwarf.  Stable astrometric measurements of GJ 867BD obtained with CTIO's 0.9-m telescope over the last decade exclude the presence of any massive planetary companions (7-18 M$_{JUP}$) with longer orbital periods (2-10 years) for the majority of orientations.  These complementary observations are also used to determine the trigonometric distance and proper motion of GJ 867BD; the measurements are consistent with the \textit{HIPPARCOS} measured values of the M2 dwarf GJ 867AC,  which is itself a 4.1 day double-lined spectroscopic binary at a projected separation of 24$\farcs$5 (216 AU) from GJ 867BD.  These new measurements strengthen the case that GJ 867AC and GJ 867BD are physically associated, making the GJ 867 system one of only four quadruple systems within 10 pc of the Sun (d$=$ 8.82 $\pm$0.08 pc) and the only among these with all M-dwarf (or cooler) components.

\end{abstract}
\section{Introduction}
Although the processes involved in single star formation have a wide variety of observational and theoretical constraints \citep[e.g.][]{2007ARA&A..45..565M}, the complex physics involved in the formation of binary and especially multiple star systems is much less understood \citep[e.g.][]{2003A&A...397..159H,Tokovinin2008, 2013arXiv1303.3028D}.  Prompt fragmentation of a collapsing prestellar core is favored as setting the initial conditions for multiple star systems \citep[e.g.][]{2002ARA&A..40..349T}, though how this depends upon the initial conditions of the cloud core (e.g. metallicity, mass, etc.) is not well understood.  Once initial fragments form, subsequent fragmentation (possibly within a disk), accretion from the remnant cloud core, N-body dynamics, and possibly orbital migration have been proposed to play competing roles in setting the final properties of multiple star systems \citep[e.g.][]{2000MNRAS.314...33B, 2001IAUS..200...23B, 2004MNRAS.351..617D, 2002A&A...384.1030S}.  While the task of identifying the relative importance of these dynamical phenomenon remains daunting, useful constraints can nevertheless be established by first determining fundamental multiple star properties, such as the overall multiplicity fraction, the distribution of mass ratios, the distribution of separations, and the mass dependence of companion properties.  As an example, one of the most well-established observational trends is that the fraction of stars in binary systems decreases toward lower stellar masses; studies of nearby stellar samples reveal that 26$\%$ of M-dwarfs have companions \citep[also see \citet{1992ApJ...396..178F, 1992ASPC...32...10H}]{2004ASPC..318..166D}, compared to 46$\%$  for Sun-like stars \citep[also see \citet{Duquennoy1991}]{Raghavan2010}.  A likely implication of this is that lower mass cloud cores fragment less often, despite any subsequent dynamical evolution.  

Formation events that lead to triple, quadruple, and even higher order stellar systems also appear to occur less often at low stellar masses.  Emerging statistics are that 2.1$\%$ of M-dwarfs occur in triple systems, while 0.03$\%$ occur in quadruple or higher order systems, compared to 5.6$\%$ of G-dwarfs occurring in triple systems and 2.2$\%$ occurring in quadruple or higher order systems \citep{Eggleton2008}.  While it is tempting to interpret this as a corresponding decrease in the frequency of multiple fragmentation events with lower cloud core mass, accurate assessments of high-order multiplicity are much more subject to observational biases and incompleteness, especially at low stellar masses.  As an example of the incompleteness that exists, \citet{Law2010} find a surprisingly high fraction of triple and quadruple systems among a targeted sample of 36 extremely wide M-dwarf pairs, formerly known only as wide M-dwarf binary stars; the results mimic the discovery that brown dwarfs are more often found to be a close binary systems when first identified as a wide companion to a star \citep{2005AJ....129.2849B}.

The most direct way to establish a more complete assessment of the multiplicity fraction over a broad range of masses is to target volume-limited samples of nearby stars.  Not only are these the closest, brightest stars with the most precisely determined kinematics, but if trigonometric distances are known, malmquist biases can be avoided.  Nevertheless, observational biases remain even for the closest stars.  Of the 348 stars known within 10 pc, 21 are G-dwarfs and 239 are M-dwarfs \citep{2006AJ....132.2360H}.  But while 2 of these 21 G-dwarfs are the primary stars in quadruple star systems  (no higher order G-dwarf systems are known), only 1 of the 239 M-dwarfs is the primary of a quintuple star system; no M-dwarf quadruple star systems are known within 10 pc (see Table~\ref{table:known})\footnote[1]{LHS1070 was claimed to be a quadruple system by \citet{1999ApJ...512..864H}, but later determined to only be a triple system \citep{2012A&A...541A..29K}.  GJ 896A and GJ 896B were on initial inspection both claimed to be spectroscopic binaries \citep{1999A&A...344..897D} because both stars showed very large variations in radial velocity.  These variations were later thought to be caused by inhomogeneous surface features combined with a large magnetic spot (priv. comm. Xavier Delfosse).}.

The deficiency of known higher order pairs in part stems from a lack of high-dispersion spectroscopy that would reveal short period companions.  Early M-dwarfs are faint in the optical and chromospherically active making spectroscopic companion searches observationally difficult.  It becomes even more challenging to search for short period companions around mid to late M-dwarfs with spectroscopy, as these stars are even fainter and tend to have higher projected rotational velocities compared with earlier M-dwarfs \citep{2009ApJ...704..975J, 2010ApJ...710..924R,  2012AJ....143...93R}.  These biases have inhibited good assessments of high order multiples.  For example, of the 90 M-dwarfs included in the spectroscopic companion search by \citet{2006ApJ...649..436E}, only 4 stars have spectral types of M4 or later.  The HARPS RV sample of 102 southern M-dwarfs only included stars with V$<$14 mag \citep{Bonfils2013}.  With the ratio of early M-dwarfs (M0-M4) to mid and late M-dwarfs (M4.5-M9) being 23/28 for stars within 5 parsecs \citep{2013AJ....146...99C}, about half of all M-dwarfs are left out of high dispersion spectroscopic surveys.    

To address the question of M-dwarf multiplicity more completely, we initiated a volume-limited, astrometric and infrared spectroscopic search for companions with orbital periods between 3 days and 8 years around mid M-dwarfs within 10 parsecs, which includes GJ 867B as a target \citep{Davison2013}.  A first result of this work is that GJ 867B is revealed to be a short-period spectroscopic binary.  We discuss the known characteristics of the GJ 867 system in \S2.  We list the astrometric results in \S3 and the photometric results in \S4, and newly assigned spectral types in \S5.  We discuss the spectroscopic results of GJ 867B and the new companion, GJ 867D, in \S6.  In \S7, we summarize the results and discuss the case for physical association of this hierarchical system.

\section{GJ 867 System}
GJ 867A (\textit{V$_{KC}$\footnote[2]{Subscripts: KC indicates Kron-Cousins (SAAO system) and used later in the text J indicates Johnson.}}=9.09 mag; \citet{1990A&AS...83..357B}) is the primary component of a widely separated visual binary.  It was classified spectroscopically by \citet{1952ApJ...116..117V} as an M type star;  we more accurately classify this star as M2.0V (see section 5).  This star is a member of young disk \citep{2007AcA....57..149K}.  Its visual companion, GJ 867B (\textit{V$_{KC}$}=11.45 mag; \citet{1990A&AS...83..357B}), is separated by 24$\farcs$5 at a position angle of 350.45$^\circ$, based on 2MASS position measurements.  The spectral type of GJ 867B is M3.5V (see section 5).   Coordinates, magnitudes and stellar properties of these stars are given in Table~\ref{table:system}.  

The two stars are presumed to be physically associated from their large ($\sim$0$\farcs$45/yr) and very similar proper motions \citep{1954AJ.....59..218D, 1995gcts.book.....V, 2011A&A...531A..92R, 2013AJ....145...44Z}.  Nevertheless, the stars have slightly but steadily changed their relative position since first recorded in 1877 with a separation of 21$\farcs$4 and a  position angle (PA) of 359$^\circ$, which presumably represents the orbital motion of the system \citep{1965ApJ...141..649H}.  

GJ 867A was discovered by \citet{1965ApJ...141..649H} to be a double-lined spectroscopic binary with an orbital period of 4.083 days.  The two components (A and C) have an estimated mass ratio (M$_C$/M$_A$) of 0.80 based on their relative velocity amplitudes.  GJ 867B was identified with high resolution optical spectroscopy as a candidate spectroscopic binary (SB) by \citet{Gizis2002}, and later was listed as either having a high rotational velocity or being a SB by \citet{Bonfils2013}. However, until now its status as a short-period binary remained unconfirmed.

\section{Astrometry}
\subsection{Observations}
Direct images of GJ 867BD\footnote[3]{When the combined light of the individual components of the binary pair cannot be separated, we give the object the name of the binary pair.} were obtained at the 0.9-m telescope at Cerro Tololo Interamerican Observatory (CTIO) from July 2003 until September 2012, yielding a total of 16 astrometric epochs over 9.17 years.  Images were obtained using the 2048 x 2048 Tektronix CCD camera on the central quarter of the CCD chip.  The CCD chip has a plate scale of 0.$\arcsec$401 pixel$^{-1}$, which gives a field of view of 6$\farcm$8 $\times$ 6$\farcm8$ \citep{2003AJ....125..332J}.  Astrometric measurements of GJ 867BD were made in the $V$$_{J}$ filter, which provided a field of five suitable reference stars.  No astrometric measurements were taken of GJ 867AC, as this star saturates before sufficient signal-to-noise ratios are obtained for the reference stars.  Each night that GJ 867BD was observed we obtained 5 to 10 images, for a total of 96 images on 16 different nights. Exposure times for an individual image varied from 25 to 100 seconds, with an average time of 54 seconds.  We aimed to observe all images within 30 minutes of the star crossing the meridian to minimize the effects of differential color refraction \citep[DCR;][]{Jao2005}.  In each image the FWHM of GJ 867BD is less than 2$\farcs$4;  this size is typically set by the seeing. The ellipticity of the target star and all reference stars are less than 20$\%$.  These criteria are adopted to eliminate elongated reference stars (e.g. binaries), and to avoid data with tracking/guiding errors.

During the time of observations, two different \textit{V${_J}$} filters were used.  The first \textit{V${_J}$} filter possessed a small crack and was replaced in February 2005 with a nearly identical \textit{V${_J}$} filter, which we confidently adopted knowing that it possessed the same transmission spectrum as the original.  After a few years of obtaining data, we noticed a few milli-arcsecond (mas) offset in the astrometric residuals of some stars that were known to be single from other techniques \citep{Subasavage2009};  we determined that the offset first appeared when we replaced the \textit{V${_J}$} filter.  The effect of using the replacement \textit{V${_J}$} filter is more clearly seen in the astrometric residuals in the right ascension axis.  After a close inspection of the first \textit{V${_J}$} filter, it was realized that the small crack does not project onto the part of the chip that is regularly used (i.e. the central quarter).  Therefore, in July 2009, we replaced the second \textit{V${_J}$} filter with the original \textit{V${_J}$} filter \citep{Riedel2010}.  All data acquired after July 2009 were with the original \textit{V${_J}$} filter.  Using data from both filters may increase the parallax error, but does not change the parallax value.  We use all points in this analysis to maximize the parallactic ellipse coverage. 

\subsection{Astrometric Reductions}
Here we briefly summarize the reduction technique; further details can be found in \citet{Jao2005} and \citet{Subasavage2009}.  We obtain photometric parallaxes using the CCD distance relations of \citet{Henry2004} from which distances are calculated for all of our reference stars.  We require reference stars to be further than 100 parsecs from the Sun, and initially assume the sums of all reference stars' proper and parallactic motions to be zero mas/yr.  We measure the centroid of our target star against the presumed fixed grid of five reference stars using the SExtractor centroiding algorithm described in \citet{1996A&AS..117..393B}.  Relative to this reference grid, we obtain the proper motion, parallax and any residual astrometric motion of the target star using the Gaussfit program from \citet{Jefferys1987}.  We use the photometric parallaxes of the reference stars from our VRI photometry and correct for the parallactic motions of our reference stars to determine the absolute parallax of GJ 867BD.

\subsection{Parallax and Proper Motion of GJ 867BD}

We measure a relative parallax for GJ 867BD of 109.45 $\pm$ 1.74 mas.  The correction for parallactic motion is 0.93 $\pm$ 0.20, yielding a final absolute parallax of 110.38 $\pm$ 1.75 mas.  This is the first parallax measurement for this star and corresponds to a trigonometic distance of 9.06 $\pm$ 0.15 parsces.  The errors on the parallax measurement represent our internal statistical precision.  To estimate our systematic errors, we determine the standard deviation of the differences between the parallaxes of previous RECONS measurements \citep[e.g.][]{Subasavage2009, Riedel2010, Jao2011} and \textit{HIPPARCOS}, which is 4.3 mas.  This error dominates our internal error and we use it here to compare our values to previously published values.  Two parallax values are available for GJ 867AC, 115.10 $\pm$ 7.40 mas \citep{1995gcts.book.....V} and 115.01 $\pm$ 1.32 mas by \textit{HIPPARCOS} \citep{2007A&A...474..653V}.  Our parallax measurement of GJ 867BD is within 1.1$\sigma$ of the weighted parallax measurement from \citet{1995gcts.book.....V} and \citet{2007A&A...474..653V} for GJ 867AC.  The similar distance measurements strengthen the case that GJ 867AC and GJ 867BD are physically associated.   The weighted mean parallax of the GJ 867 system is 113.37 $\pm$ 1.04 mas, which corresponds to a distance of 8.82 $\pm$ 0.08 parsces.

Our measurements of the proper motion amplitude, $\mu$, and the proper motion position angle for GJ 867BD are 421.8 $\pm$ 0.6 mas/yr and 97.1 $\pm$ 0.1 degrees, respectively.  The errors on the proper motion amplitude represent our internal statistical precision.  To estimate our systematic errors, we again use the standard deviation of the differences between the proper motion amplitudes of previous RECONS measurements \citep[e.g.][]{Subasavage2009, Riedel2010, Jao2011} and \textit{HIPPARCOS}, which is 16.8 mas/yr.  Similar to our parallax errors, the systematic error for the proper motion amplitude dominates the error budget.  \citet{2003ApJ...582.1011S} measure the $\mu$ of GJ 867BD to be 431.2 $\pm$ 10.2 mas/yr and the proper motion position angle to be 98.1 degrees, which is within 0.6$\sigma$ of our measurement of the $\mu$ of GJ 867BD; they do not provide an uncertainty on their measured position angle.  GJ 867AC has a $\mu$ of 455.8 $\pm$ 2.7 mas/yr and a position angle of 99.9 degrees \citep{2003ApJ...582.1011S}.  The difference in proper motion amplitudes for the AC and BD components is $\sim$35 mas/yr, which is slightly larger than our systematic error of 16.8 mas/yr and nevertheless consistent with orbital motion. Table~\ref{table:motion} summarizes previous parallax and proper motion measurements.  

\subsection{Detection Limits for Astrometric Companions}
The precise position measurements of GJ 867BD spanning almost 10 years allow us to set mass limits on long period astrometric companions.  This first requires an accurate assessment of the position uncertainties which we estimate empirically.  This is done by characterizing the statistical uncertainty associated with each epoch's position measurement.  To do this, we calculate the mean position error per night, which is the average standard deviation of position offsets, once parallactic and proper motions are removed.  The mean position errors for all nights in RA and DEC are 2.31 and 4.00 mas, respectively.  The average residual deviation from the position predicted by the proper motion and parallax fit is calculated as the standard deviation of the nightly offsets, and in RA and DEC are 1.37 and 3.83 mas, respectively.  As we are measuring the predicted position with better precision than the average error, this may be an indicator that we are slighty over-estimating the mean position errors.

In order to set limits on the presence of companions orbiting GJ 867BD with orbital periods between 2 to 10 years, we perform 100,000 Monte Carlo simulations with a range of orbital properties.  Orbital periods are restricted to 2, 4, 6, 8, and 10 years, which are representative of our observational sampling and baseline.  We estimate the mass of the primary to be 0.29 $\pm$ 0.06 M$_{\odot}$ based on the mass luminosity relation of \citet{1993AJ....106..773H}.  Although we adopt this mass for all of our calculations, we note that this mass is based on the brightness of both GJ 867B and GJ 867D, as we are unable to deconvolve the binary and determine magnitudes and colors estimates for each component.  The companion star GJ 867D appears to contribute little light to the system (see section 6.3), and thus is unlikely to severly bias this estimate.  This mass determinaton is slightly higher than the mass deterimed by the spectral type of GJ 867B and the spectral type mass relations in \citet{Kraus2007} of $\sim$0.25 M$_{\odot}$, but consistent within the errors.  We use the weighted distance measurement of the GJ 867 system reported in section 3.3.  We allow the inclination to vary between 0 and 90 degrees and the eccentricity to vary between 0 and 0.9.  We allow the longitude of periastron, $\omega$, and the longitude of the ascending node, $\Omega$, to vary from 0 to 180 degrees.  This simulation suggests that 90$\%$ of the time, we would have been able to detect, with 99.7$\%$ (or 3$\sigma$) confidence, a companion with a mass of at least 18 M$_{JUP}$ in a 2 year period, 11.5 M$_{JUP}$ in a 4 year period, 9 M$_{JUP}$ in a 6 year period, 7.5 M$_{JUP}$ in a 8 year period and 7 M$_{JUP}$ in a 10 year period.  The implication is that GJ 867BD is unlikely to have any brown dwarf or massive planetary companions with periods between 2 to 10 years.  The orbital periods correspond to separations of 1.1 to 3.1 AU (0$\farcs$07 to 0$\farcs$22, if face on).

\section{New System Photometry}
\subsection{Photomertic Observations}
We obtained photometric measurements of GJ 867BD and its reference stars on three different nights using an identical setup as that used for the astrometric measurements at the CTIO 0.9-m telescope.  As our initial goal was to obtain photometry of GJ 867BD, we allowed GJ 867AC to saturate in our images, and therefore provide no new apparent magnitude measurements for GJ 867AC.  We obtained \textit{V${_J}$}, \textit{R$_{KC}$}, and \textit{I$_{KC}$} photometry of GJ 867BD that were calibrated using at least 10 photometric standard stars per night from \citet{Landolt1992} and \citet{Graham1982}.  All images were obtained at an airmass of less than 2.  

\subsection{Photometric Reductions}
Observations of the photometric standards are used to determine transformation equations and to create the extinction curves, enabling us to calculate absolute photometry for GJ 867BD.  Specific details are given in \citet{Jao2005} and \citet{Winters2011}.  The apparent magnitudes through the \textit{V${_J}$}, \textit{R$_{KC}$}, and \textit{I$_{KC}$} filters are 11.47 mag, 10.29 mag, 8.77 mag, respectively, with an error of 0.01 mag each.  Our measurements are within 0.02 mag of the measured photometry by \citet{1990A&AS...83..357B}.  

Although astrometric observations do not allow absolute photometry, relative photometry can be extracted.  Consequently, all images used to search for variability are in the V$_J$ filter.  We measure the standard deviation of the instrumental magnitude of GJ 867BD, ($\sigma$$_{mag}$), relative to reference stars for over 16 epochs spanning 9.17 years to assess the amount of photometric variability.  For GJ 867BD, we use the same set of reference stars in the astrometric reduction. The average $\sigma$$_{mag}$ is 32.4 millimags, which is above the mean activity level of most field dwarfs \citep[$\sim$13 millimags;][]{Jao2011}, but is not surprising given its classifcation as a flare star \citep{1978IBVS.1407....1B}. 

\section{New Spectral Types}
New spectral types are determined using the RECONS classification system, which focuses on K and M dwarfs.  The spectral region used for this covers 6000$-$9500\AA~at a resolution of 8.6\AA; further details are given in \citet{2008AJ....136..840J} and \citet{Riedel}.  Observations were made using the R-C Spectrograph on the CTIO 1.5m with the Loral 1200 $\times$ 800 CCD camera, utilizing the \#32 grating in first order with tilt 15.1$^\circ$, and order blocking filter OG570.  We remove the H$\alpha$ feature and the telluric features on the spectra using the sky transmission map from \citet{Hinkle2003}.  We then compare our stellar spectra to our library of standards spectra from \citet{2002AJ....123.2002H} and determine that the best match with the lowest standard deviation to be the spectral type of our star.   Spectral types are accurate to 0.5 spectral sub-classes.  We determine the spectral types of GJ 867AC and GJ 867BD based on the combined light of the spatially unresolved pairs to be M2.0 and M3.5, respectively.  These values are consistent to within 0.5 sub-classes of previously adopted values \citep[e.g.][]{2004AJ....128..463R}.

\section{High Dispersion Spectroscopy}
\subsection{Observations}
Spectrographic observations were made at NASA's Infrared Telescope Facility, a 3 meter telescope, using the high resolution infrared echelle spectrograph, CSHELL \citep{1990SPIE.1235..131T, 1993SPIE.1946..313G}.  Spectra were centered at 2.298 microns (vacuum) and span approximately 50 \AA.  This wavelength region was chosen to  obtain precise RVs by measuring the positions of the rich photospheric $^{12}$CO R lines (2.3 microns) relative to a wavelength solution set by telluric methane absorption features in the Earth's atmosphere \citep[e.g.][]{Blake2010, Bailey2012}.  Observations were made using the 0$\farcs$5 slit, yielding a measured resolving power of $\sim$ 46,000.  Ten measurements of GJ 867B were obtained spanning a temporal baseline of 100 days between August 22, 2012 and November 29, 2012.   Each measurement consists of two spectra taken in quick succession of one another along the vertical direction of the slit, separated by 10$\arcsec$.  The two different positions are referred to as nod positions.  The first three RV measurements were obtained during late August 2012, over a 4 day period.  At that time, we noticed a large RV variation of almost 18 km s$^{-1}$ and flagged this variable candidate as a high priority target for our upcoming runs to confirm it as a spectroscopic binary.  Exposure times averaging $\sim$500 seconds each were set to obtain predicted signal-to-noise ratios (SNRs) of $\sim$100 per exposure, and resulted in pair combined SNRs ranging from 70 to 230.  Along with observations of GJ 867B, spectra of either the A7V star Altair or the A1V star Sirius were obtained to gain an initial estimate of the wavelength solution and the instrumental profile of CSHELL, since these spectra contain only telluric features.  Finally, 20 flats and darks with integration times of 10 or 20 seconds were taken each night.
\subsection{Spectroscopic Reductions}
A brief overview highlighting the aspects of our spectral extraction and radial velocity (RV) fitting routine will be given here.  For a full explanation of the data reduction technique, the reader is referred to \citet{Bailey2012} and \citet{Davison2013}.  

To prepare the 2D images for spectral extraction, we first correct for flat fielding using dark-subtracted, median-combined flat field images.  Then, we remove sky emission, detector bias and dark current from the spectral images by subtracting each individual image from its pair image taken at a different nod position that immediately preceded or followed the image.  We assume that the changes in the detector and spectrograph properties are negligible over the short period in which the two spectra were taken at the different nod positions.  Then, we perform an optimal extraction of the spectra as implemented for nod subtracted pairs \citep{1986PASP...98..609H, Bailey2012, Tanner2012}.  Unlike a standard extraction that simply sums over all the pixels to determine the 1D spectral profile, an optimal extraction sums pixels weighted by a variance image of the spectrum's spatial profile to determine the spectral profile.  It is assumed that the spectral profile varies smoothly in the direction parallel to the dispersion.   Using a weighted image to determine the spectral profile gives the advantage that it minimizes the noisy contribution of the profile wings and eliminates the majority of the features on the spectral profile caused by cosmic rays or hot pixels.

\subsection{Spectroscopic Analysis and Modeling as a Single-lined Spectroscopic Binary}
If GJ 867BD is a double-lined spectroscopic binary, we would expect it to appear double-lined in at least some of our spectra.  Visual inspection showed no sign of a secondary component.  Likewise, if its companion is of comparable brightness, we might observe broader absorption lines when the stars in the system are at opposition.  Broader absorption lines would then be translated by our analysis code into larger $v$sin$i$ values.  However, all 10 $v$sin$i$ measurements are consistent to within 3.1 km s$^{-1}$ and show no apparent correlated variation with the changes in RV.  These factors lead us to believe that this star is a single-lined spectroscopic binary (SB1).  Also, there is a lack of obvious remaining residuals in the single star spectral modeling, described below.  

We subsequently analyze the system as an SB1.  After extracting the spectra, we fit each observation to high resolution spectral models that are convolved to the resolution of CSHELL.  The models are formed by combining synthetic spectra created from NextGen \citep{1999ApJ...512..377H} with empirical telluric models extracted from ultra high resolution KPNO/FTS telluric spectra of the Sun \citep{1991aass.book.....L}.  Our synthetic model is composed of spectral features from a single M3.5V star.  In order to make a realistic model, we assume the starting stellar parameters of GJ 867B as follows:  We estimate the stellar temperature to be 3200 K based on its spectral type of M3.5V, using the relationship from \citet{Kraus2007};  we set its surface gravity (log(g)) to be 4.8 (cgs), as this number is consistent with other field dwarfs of similar mass \citep{Hillenbrand2004}.  Our model spectrum consists of 19 parameters.  The limb darkening coefficient parameter is set to 0.6, a reasonable assumption for cool stars at infrared wavelengths \citep{Claret2000} and the remaining 18 parameters are left to vary.  Of those variable parameters, 3 of those parameters are used to solve the wavelength solution and 9 of those parameters are Gaussians used to model the PSF of the spectrum.  The remaining 6 are the depth of the telluric spectra, depth of the synthetic spectra, $v$sin$i$, RV, and 2 continuum normalization constants.  

To determine the 12 parameters related to the wavelength and instrumental profile, we first fit the empirical telluric spectrum to A star observations.  We obtain the optimal parameters by minimizing the variance weighted reduced $\chi$$^{2}$.  On the one night when an A star was not observed, we averaged parameters determined on nights with A star observations and during the same run together to estimate initial guesses for the wavelength solution and instrumental profile.  This step is critical, as our optimization routine is sensitive to initial guesses.  

\begin{figure}
\begin{center}
\includegraphics[scale=0.32,angle=90]{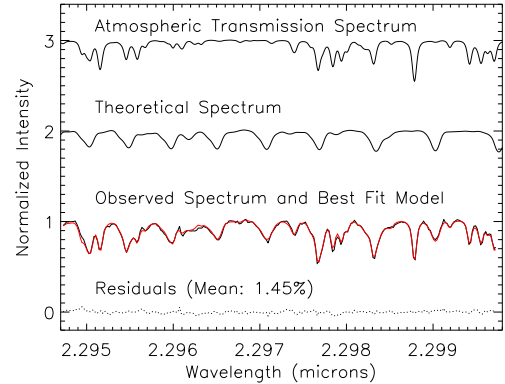}
\caption[fig1] {Spectra of GJ 867BD.  Spectra are modeled by combining a telluric spectrum (top spectrum), which provides an absolute wavelength reference, with a theoretical stellar spectrum (2$^{nd}$ spectrum).  The CSHELL spectrum of GJ 867B is shown (black; 3$^{rd}$ spectrum) in comparison with the the best fit (red; 3$^{rd}$ spectrum).  The mean residuals of the best fit spectrum are 1.45$\%$ (bottom spectrum).  The average mean residuals for all CSHELL spectra obtained is 3.09$\%$.}
\label{fig:spectra}
\end{center}
\end{figure}

After obtaining these initial parameters, we fit the combined model of the telluric and synthetic spectra to our observed CSHELL spectra.  During this iterative process, we first fit for the wavelength solution, the depth of the telluric and synthetic spectra, the RV, and the two normalization constants.  Then, on our second iteration, we fit for $v$sin$i$, the depth of the telluric and model spectral features and the two normalization constants.  On our final iteration, we optimize the wavelength solution, the depth of the telluric and synthetic spectra, the RV, and the two normalization constants.  All of our fits are optimized using AMOEBA, which uses a downhill simplex method to find the best solution for multiple variables \citep{Nelder1965}.  We restrict AMOEBA to specified ranges of physically reasonable solutions.  An example of an optimally extracted spectrum of GJ 867BD, the telluric model, the stellar model and the best fit combination are shown in Figure~\ref{fig:spectra}.


\subsection{Spectroscopic Results}
The determined RVs and projected rotational velocities ($v$sin$i$ values) are listed in Table~\ref{table:rv}.  The observed RV error is a combination of theoretical photon noise error, intrinsic stellar error and instrumental error ($\sigma$$_{obs}$$^{2}$ $=$ $\sigma$$_{photon}$$^{2}$ $+$ $\sigma$$_{stellar}$$^{2}$  $+$ $\sigma$$_{instr}$$^{2}$).  The photon error is calculated based on the prescription by \citet{1996PASP..108..500B}.  We assume the intrinsic stellar error to be zero.  We know this is an underestimate of the stellar error for GJ 867B, which is an active flare star.  However, we do not have enough data to accurately assess the error caused by intrinsic stellar error for this star.   We adopt the average instrumental error of  68 m s$^{-1}$ determined for 10 stars with low rotational velocities from our survey of mid M-dwarfs \citep{Davison2013} as the instrumental error for our RV measurements of GJ 867BD.  The average $v$sin$i$ value from our fitting prescription is 8.74 $\pm$ 0.98 km s$^{-1}$.  The error on the $v$sin$i$ measurement is set from the standard deviation of the independent measurements from all 10 epochs.

\subsection{Periodicity Analysis}
\subsubsection{The Orbital Period}
\begin{figure}
\begin{center}
\includegraphics[scale=0.5,angle=0]{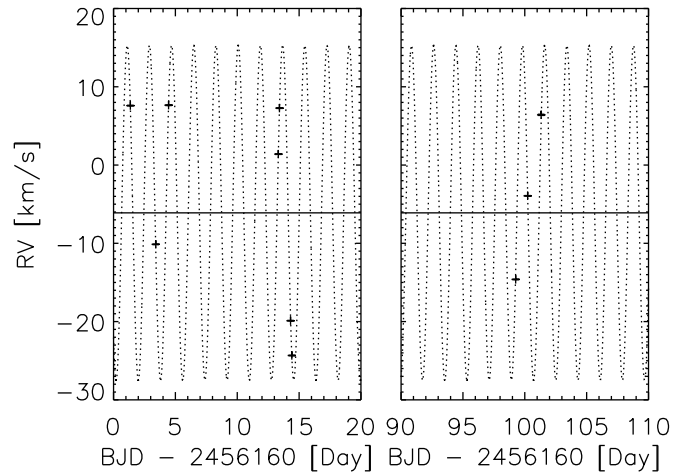}

\caption[fig4] {RV measurements of GJ 867B versus barycentric Julian date. A Keplerian curve is fit to the RV measurements, centered on the systemic velocity of -6.1 km s$^{-1}$.  The errors on the RV meausurements are smaller than the plot symbols.}
\label{fig:rvcurve}
\end{center}
\end{figure}

\begin{figure}
\begin{center}
\includegraphics[scale=0.44,angle=0]{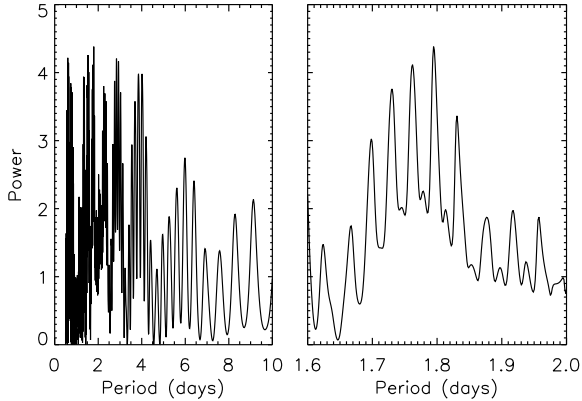}

\caption[fig4] {Periodograms of the RV measurements of GJ 867BD. The \textit{left panel} illustrates the periodogram for the full range of periods considered, while the \textit{right panel} is the portion containing the most significant period of 1.795 $\pm$ 0.017 days.}
\label{fig:period}
\end{center}
\end{figure}

\begin{figure}
\begin{center}

                \includegraphics[scale=.44,angle=0]{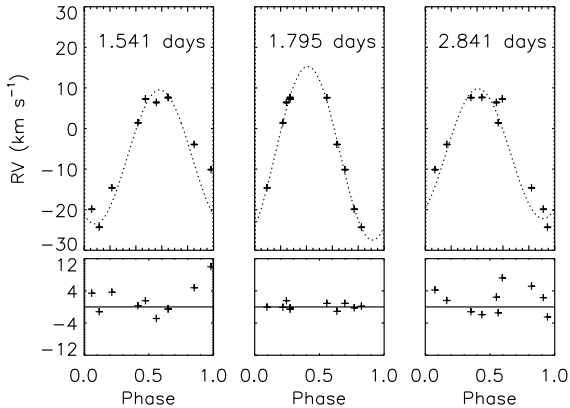}

                          \caption{Phased RV Curves of GJ 867BD.  Shown are the RV curves of GJ 867BD phased to 1.541, 1.795, and 2.841 days, which are the three periods with the highest significance from the periodogram analysis.  The black dotted line is the best fit Keplerian model; the best fit residuals are shown in the bottom panels.  The period of 1.795 days is strongly favored because of its much smaller residuals.}

\label{fig:oc}

\end{center}
\end{figure}

GJ 867B shows a large change in RV on timescales of less than one day (Figure~\ref{fig:rvcurve}).  To determine if the changes in RV are periodic, we use a Lomb Scargle algorithm to search for periods between 0.5 and 10 days.  The highest power of the periodogram favors a period of 1.795 days.  Unfortunately, because of the limited sampling, there are many other peaks of comparable power (Figure~\ref{fig:period}).  The highest peak at 1.795 days has a power value of 4.37, which is closely followed by power values of 4.24 and 4.21 for the next most likely peaks at periods of 1.541 and 2.841 days, respectively.  To confirm that the highest peak is representative of the best-fit period, we conduct an independent $\chi$$^{2}$ analysis, where we fit circular orbits to the top three periods given from the periodogram fit.  Knowing the orbital period is less than 3 days, we assume the orbits are circularized due to tidal effects \citep{1984IAUS..105..411M}.  We optimize the circular fitted curves to give the lowest observed-minus-calculated (O-C) RV value using the AMOEBA minimization routine (Section 3.4); the results are shown in Figure~\ref{fig:oc}.  The average (O-C) value for 1.795 days is 0.58 km s$^{-1}$ and is almost 5 times smaller than the (O-C) values for the periods of 2.841 days (2.88 km s$^{-1}$) and 1.541 days (3.00 km s$^{-1}$).  Similarly, the $\chi$$^{2}$ value for 1.795 days is 27 and 38 times smaller than the $\chi$$^{2}$ values at 1.54 days and 2.841 days, respectively.  From these comparisons, we conclude that 1.795 days is the orbital period of the BD system.  

\subsubsection{The Uncertainties in the Orbit}
To determine the orbital properties and the associated uncertainties of the orbit, we use the non-linear least squares fitting IDL routine, mpfitfun.pro \citep{2009ASPC..411..251M} to model our data as a sinusoidal curve.  The amplitude of the fitted curve is 21.4 $\pm$ 0.5 km s$^{-1}$, the velocity of the system is $-$6.1 $\pm$ 0.3 km s$^{-1}$, and the epoch of conjunction is JD 2456160.5611 $\pm$ 0.0268.

The peak with the most likely period is in a cluster of aliased peaks.  To test whether the neighboring peaks represent reasonable orbital solutions, we again fit circular orbits to the top three aliasing peaks in closest proximity to the favorable period.  The (O-C) values for these peaks range from 3.16 km s$^{-1}$ to 5.56 km s$^{-1}$ making these values significantly larger than the (O-C) values from the top three likely periods given by the periodogram fit.  We conclude that neither of these periods represent good fits to our data and determine the uncertainty in the period from the FWHM of the central peak only, which yields a final best fit period and uncertainty of 1.795 $\pm$ 0.017 days.  

The average difference between the (O-C) value for our best fit circular orbit of 0.58 km s$^{-1}$ is significantly higher than our average estimated error for this star based on the theoretical photon noise error and the instrumental error.  This could be caused by a few possibilities.  GJ 867B could have a slightly elliptical orbit, and therefore the deviation we see from the observed compared to calculated value is real.  Alternatively, our estimated errors could be low, as we did not account for any intrinsic stellar error.  GJ 867B is a flare star with a nominally higher rotational velocity, which could be an indicator of a higher amount of intrinsic stellar jitter compared to other older field dwarfs.  We caution that if the RV errors are too low, then the calculated errors on the orbital properties will be underestimated as well.

\subsection{Limits on the Mass of GJ 867D}
Without a direct detection of GJ 867D, we can only provide a lower mass limit based on the observed reflex motion of GJ 867B using Kepler's third law.  We assume the orbit is circular and that the mass of the primary star is 0.29 $\pm$ 0.06 M$_{\odot}$ (Section 3.4). Under these assumptions the minimum mass of GJ 867D is 61 $\pm$ 7 M$_{JUP}$ (0.059 $\pm$ 0.007 M$_{\odot}$); the uncertainty in the minimum mass is determined primarily by the uncertainty in the primary star's mass.  GJ 867D could be a brown dwarf.  The spectral properties of GJ 867AC and GJ 867BD are summarized in Table~\ref{table:mass}.   

\begin{figure}
\begin{center}
\includegraphics[scale=.44,angle=0]{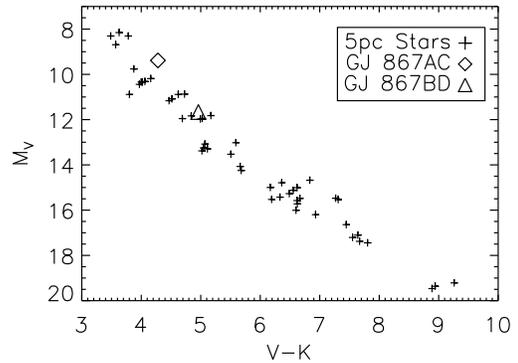}
\caption[fig4] {Absolute V Magnitude vs. (V-K) Colors.  This color-absolute magnitude diagram shows GJ 867AC and GJ 867BD relative to all known M-dwarfs within 5 parsecs.  This plot shows that GJ 867BD is not significantly over-luminous relative to this empirical main sequence, consistent with GJ 867BD being a single-lined SB and GJ 867D being low mass.}F
\label{fig:color}
\end{center}
\end{figure}

The brightness of GJ 867BD supports the conclusion that GJ 867D is very low mass.  We compare the  absolute magnitude in V to the (V-K) color of stars within 5 parsecs \citep{2013AJ....146...99C} to GJ 867 system (see Figure~\ref{fig:color}).  GJ 867BD lies within the broad band making up the main sequence.  Therefore, there is no evidence that the companion contributes much light, consistent with its undetected spectral features.
\section{Summary}
We report infrared spectroscopic observations obtained with the CSHELL spectrograph that reveal GJ 867B to be a single-lined SB with an orbital period of 1.795 $\pm$ 0.017 days; the components are labeled B and D.  Assuming that GJ 867B has a mass of 0.29 M$_\odot$, then GJ 867D has a minimum mass ($m$sin$i$) of 61 $\pm$ 7 M$_{JUP}$; it could be a brown dwarf.  Astrometric measurments spanning nearly 10 years are used to determine  the trigonometric parallax and proper motion of GJ 867BD.  The trigonometric parallax of GJ 867BD is consistent to within 1.1$\sigma$ of the measured parallax of GJ 867AC by \textit{HIPPARCOS}, confirming that the components of this $24\farcs5$ pair are at similar distances.  The measured proper motion amplitude of GJ 867BD is 421.8 $\pm$ 0.6 mas/yr, which is similar to but  slightly smaller than the proper motion amplitude of GJ 867AC \citep[455.8 $\pm$ 2.7 mas/yr,][]{2003ApJ...582.1011S}; both stars are moving in the same direction to within 2 degrees.  The difference (-34.0 mas/yr) can easily be explained by orbital motion.  Adopting the minimum masses in Table 5 and assuming the projected separation (216 AU) is the semi-major axis of the AC and BD pair, these components would be moving relative to one another at 1.8 km/s, corresponding to 43.2 mas/yr (at 8.82 pc) if in a face-on circular orbit.  If in a somewhat inclinded orbit (and if the components are more massive than the minimum masses used here), orbital motion could also explain the difference in the systemic velocities of GJ 867BD, at $-6.1\pm 0.3$ km/s (Table 5), and GJ 867AC, at $-8.7$ km/s \citep{1965ApJ...141..649H}; we consider this apparent discrepancy less relevant, however, since the  systemic velocity for GJ 867AC is based on much coarser precision individual measurements (1.4 km/s) and no formal uncertainty on the systemic velocity is provided.  Altogether, the similar proximity, distance, and space motion of these two high proper motion spectroscopic binaries establish them as very likely physically associated.  As such, the 4 star system GJ 867 becomes one of only four quadruple systems within 10 pc, and the closest quadruple  system to the Sun with an M dwarf primary.

The discovery of this nearby quadruple system suggests that many more hierarchical low mass systems may exist nearby, but have not been found because of observational biases.  As highlighted in the introduction, the discovery and characterization of systems like this  potentially offer a powerful tool for constraining the uncertain physics involved in binary and multiple star formation.  We note that nearby systems like GJ 867 are especially valuable, since these stars are close enough to have the spectroscopic orbits that can be spatially resolved interferometrically, allowing accurate masses of these systems to be determined.

\acknowledgements  
The authors would like to thank Guillem Anglada-Escud{\'e}, Doug Gies, John Lurie, and Robert Parks for intellectual conversations and Travis Barman for generating our synthetic spectral models from NextGen \citep{1999ApJ...512..377H}.  The authors would like to note that the astrometric data were obtained by the RECONS team through the NOAO Surveys Program and the SMARTS consortium.  The infrared spectrocopic data were obtained from the Infrared Telescope Facility, which is operated by the University of Hawaii under Cooperative Agreement no. NNX-08AE38A with the National Aeronautics and Space Administration, Science Mission Directorate, Planetary Astronomy Program.  The authors are grateful for the telescope time granted to them.  This work was supported by the National Science Foundation through a Graduate Research Fellowship to Cassy Davison.  This project was funded in part by NSF/AAG grant no. 0908018 and the NSF grant no. AST05-07711 and AST09-08402.


\normalsize

\onecolumn
\begin{table}[!htp]
\caption{Quadruple or Higher Order Systems within 10 pc}
\smallskip
{\small
\begin{tabular}{cccc}
\tableline							  
\tableline							  
Primary     & Spectral Type       & Parallax               &  Configuration$^c$  \\	
Star        & of Primary$^a$     & (mas)$^b$           &                      \\	  
\tableline
GJ 423A    &   G0VJ$^{1}$    & 119.51$\pm$0.79$^{1,2,3}$         & AC-BD              \\	
GJ 570A    &   K4V$^{2}$     & 170.62$\pm$0.67$^{1,2,3}$         & A-BC-D        \\	
GJ 695A    &   G5IVJ$^{3}$   & 120.32$\pm$0.16$^{2,3}$         & AD-BC            \\	 
GJ 867A    &   M2VJ$^{4}$    & 115.01$\pm$1.30$^{2,3}$         & AC-BD          \\
GJ 644A    &   M3.5VJ$^{4}$  & 154.96$\pm$0.52$^{1,2,3,4}$         & ABC-D-GJ 643 \\
\tableline

\tableline				      

\end{tabular}
\label{table:known}
}
\small
\bigskip
\newline

(a) The letter "J" indicates a spectral type based on light from a spatially unresolved pair.  References. (1) \citet{1989PDAO...17....1B}; (2) \citet{2006AJ....132..161G}; (3) \citet{1991adc..rept.....G}); (4) this work.  
\newline
(b) Parallaxes are weighted parallaxes from the following: (1) \citet{1999A&A...341..121S}; (2) \citet{1995gcts.book.....V};  (3) \citet{2007A&A...474..653V};  (4) \citet{2005AJ....130..337C};  (5) \citet{1998A&AS..133..149M};  (6) \citet{1998A&A...330..585M}.
\newline
(c) Widely separated pairs are indicated with a hypen.

\end{table}


\begin{table}[!htp]
\caption{Characteristics of the GJ 867 System}
\smallskip
{\small
\begin{tabular}{ccccccccccc}
\tableline							  
\tableline							  
Star & R.A.         & Decl.        & V$_{KC}$ & R$_{KC}$ & I$_{KC}$ & J       & H       & K$_{S}$       & Spectral & $v$sin$i$ \\	  
     &(J2000.0)     &(J2000.0)     &   (mag)  &   (mag)  &   (mag)  &   (mag) &  (mag)  &   (mag)       &   Class                & (km s$^{-1}$) \\	 	  
\tableline                                                          						       
GJ 867AC & 22 38 45.60$^{1}$ & -20 37 16.1$^{1}$ & 9.09$^{2}$     &  8.08$^{2}$   &  6.88$^{2}$    &5.67$^{1}$ & 5.11$^{1}$ & 4.80$^{1}$ & M2.0VJ$^{3}$     &   4.7$^{4}$ \\	  			       
GJ 867BD & 22 38 45.31$^{1}$ & -20 36 51.9$^{1}$ & 11.47$^{6}$    & 10.29$^{6}$   &  8.77$^{6}$    &7.34$^{1}$ & 6.82$^{1}$ & 6.49$^{1}$ & M3.5VJ$^{3}$     &   8.7$^{3}$\\	   			      
					       
\tableline				      

\label{table:system}
\end{tabular}
}
\small
\bigskip
References. (1) \citet{2003yCat.2246....0C}; (2) \citet{1990A&AS...83..357B}; (3) this work; (4) \citet{Houdebine2010}.  

\end{table}



\begin{table}[!htp]
\caption{Parallax and Space Motions of the GJ 867 System}
\smallskip
{\small
\begin{tabular}{ccccccc}
\tableline	
\tableline						  
Group  &Targets & abs. $\pi$ &   $ \mu$       & P.A.  & V$_{tan}$     & reference  \\	  
       &        &    (mas)   &  (mas/yr)      & (deg) & (km s$^{-1}$) &         \\	 	  
\tableline                                                          						       
RECONS &GJ 867AC &  ...                  &     ...                 &  ...                  & ...  & ...  \\	   		
RECONS &GJ 867BD &110.38$\pm$1.75 &  421.8$\pm$0.6 &   97.1$\pm$0.1   & 18.1 & this paper \\	   			      
\tableline
HIP    &GJ 867AC & 115.01$\pm$1.32 &  455.9$\pm$1.7         &    100.0       &   18.8   & {\citet{2007A&A...474..653V}}   \\	  			       
HIP    &GJ 867BD &   ...                    &        ...                      &   ...           & ...     & ...    \\
\tableline					       
NLTT   &GJ 867AC &      ...                 &   455.8$\pm$2.7        &  99.9         &   ...  &  {\citet{2003ApJ...582.1011S}}\\	  			       
NLTT   &GJ 867BD &      ...                 &   431.2$\pm$10.2        &  98.1            & ...   &  {\citet{2003ApJ...582.1011S}}       \\
\tableline				      
YPC    &GJ 867AC & 115.10$\pm$7.40 &   459        &  97          &  18.9      & {\citet{1995gcts.book.....V}} \\	  			       
YPC    &GJ 867BD & ...                      &   449        &  99          &    ... & {\citet{1995gcts.book.....V}}     \\
\tableline
LEP    &GJ 867AC   & ...   &   458.0      &  100       &   ...     & {\citet{2011AJ....142..138L}}  \\
LEP    &GJ 867BD   &  ...  &     ...      &    ...       &   ...    & {\citet{2011AJ....142..138L}}   \\
\tableline
TYCHO-2    &GJ 867AC & ...   &  455.8$\pm$2.7       &  99.9      &   ...  & {\citet{2000A&A...355L..27H}}     \\
TYCHO-2    &GJ 867BD &...    &  ...         &   ...      & ...  & {\citet{2000A&A...355L..27H}}     \\
\tableline
UCAC4      &GJ 867AC & ...   &  457.6$\pm$11.3       &   100.0      &   ...  & {\citet{2013AJ....145...44Z}}     \\
UCAC4      &GJ 867BD & ...   &  432.6$\pm$15.2       &   100.6      &   ...  & {\citet{2013AJ....145...44Z}}      \\
\tableline
PPMXL     &GJ 867AC &  ...   &   454.9$\pm$2.7      &   100.2      &   ...  & {\citet{2011A&A...531A..92R}}     \\
PPMXL     &GJ 867BD & ...    &   419.1$\pm$4.6      &   98.4     &  ...     & {\citet{2011A&A...531A..92R}}   \\
\tableline
Weighted Mean & GJ 867  & 113.37$\pm$1.04  &   426.2$\pm$0.4      &     99.2 & 18.6  &\\
\tableline    
\label{table:motion}

\end{tabular}
}
\small
\newline

Notes. Several numbers related to characterizing the motions of this system are given without errors.  All parallax and proper motion values given with errors of GJ 867AC and GJ 867BD are included in the calculation for the weighted mean values of GJ 867.  The weighted mean position angle and weighted mean tangential velocities values are pure averages of the available numbers, as none of these numbers have listed errors.
\end{table}

\begin{table}[!htp]
\caption{Spectroscopic Results for GJ 867BD}
\smallskip
{\small
\begin{tabular}{ccccc}
\tableline							  
\tableline							  
 JD-240000  &   Orbital & Radial Velocity & SNR           & $v$sin$i$\\	  
             &  Phase  & (m s$^{-1}$)    &               & (km s$^{-1}$)\\	  
\tableline                                                          						       
56161.36&     1.00        &    7594	$\pm$49&	177&	9.3\\
56163.41&     1.25        &	-10132	$\pm$53&	168&	8.4\\
56164.45&     0.49        &    7645	$\pm$61&	135&	9.8\\
56173.32&     0.40        &	1396	$\pm$53&	162&	7.6\\
56173.41&     0.49        &	7274	$\pm$62&	138&	8.9\\
56174.31&     1.39        &   -19888	$\pm$46&	204&	9.3\\
56174.41&     1.48        &   -24330	$\pm$44&	228&	8.4\\
56259.26&     0.22        &	-14613  $\pm$47&	127&	9.5\\
56260.23&     1.20        &	-3936	$\pm$41&	197&	9.5\\
56261.32&     0.50 	  &    6415     $\pm$139&	71&	6.7\\
					     					      
\tableline				      

\end{tabular}
\small
\bigskip
\newline

Notes. The reported RV measurements have been corrected for the barycentric motion of Earth.

\label{table:rv}

}
\end{table}

\begin{table}
\caption{Characteristics of the Spectroscopic Binaries}
\smallskip
{\small
\begin{tabular}{ccc}
\tableline							  
\tableline							  

                       & GJ 867AC$^{a}$                      & GJ 867BD \\
\tableline   
Period (days)          &  4.08322 $\pm$ 0.00004        &  1.795 $\pm$ 0.017       \\
$\gamma$ (km s$^{-1}$) & $-$ 8.7		       &  $-$6.1 $\pm$ 0.3            \\
K$_1$	(km s$^{-1}$)  & 46.8 $\pm$ 0.45               &  21.4 $\pm$ 0.5           \\
K$_2$	(km s$^{-1}$)  & 58.1 $\pm$ 0.46               &   ...                     \\
T$_0$   (days)         & JD 2437144 $\pm$ 0.006        & JD 2456160.5611 $\pm$ 0.0268                       \\
M$_1$ (M$_{\odot}$)    &  0.271 $\pm$ 0.006$/$sin$^3$$i$      &   0.29 $\pm$ 0.59$^{b}$        \\
M$_2$ (M$_{\odot}$)    &  0.218 $\pm$ 0.005$/$sin$^3$$i$   & $\ge$ 0.056 $\pm$ 0.007$^{b}$     \\
a$_1$sin$i$ (km)       &  2.63 ($\pm$ 0.03) x 10$^6$     &  0.528  ($\pm$ 0.02) x 10$^6$ \\
a$_2$sin$i$ (km)       &  3.26 ($\pm$ 0.03) x 10$^6$     &    ...                   \\
e                      &  0.010 $\pm$ 0.010            & 0$^{c}$                        \\
$\omega$                     &  356$^o$ $\pm$ 32$^o$         & 0$^{c}$                        \\
\tableline				      
\label{table:mass}

\end{tabular}
}
\small
\bigskip

(a) Binary properties for GJ 867AC are from \citet{1965ApJ...141..649H}.
\newline
(b) The mass estimate of the primary is based on a mass luminosity relation while that of the secondary is a minimum mass ($m$sin$i$; see text).  
\newline
(c) The eccentricity and longitude of periastron are assumed to be zero.

\end{table}

\twocolumn

\label{fig:GJ0867B_AST.ps}
\begin{figure}
\begin{center}
\includegraphics[scale=.32,angle=90]{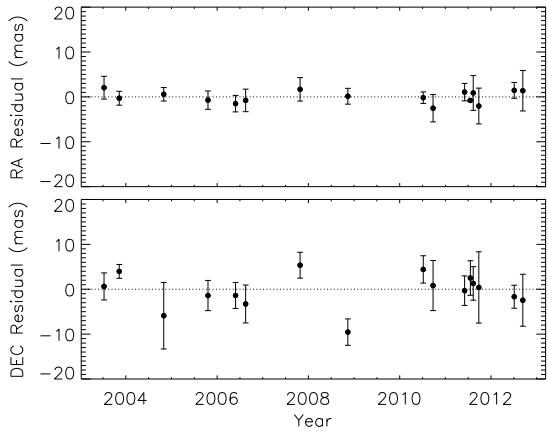}
\caption[fig4] {Astrometric residuals of GJ 867BD after solving for the proper motion and parallactic motion.  The mean position errors in RA and DEC are 2.31 mas and 4.00 mas, respectively.  These are slightly larger than the average residual deviations about zero, which are 1.37 mas and 3.83 mas for RA and DEC, respectively.  From these data we see no indication of long term variations and exclude the presence of most brown dwarf and massive planets with long orbital periods (2-10 years).}
\label{fig:AST}
\end{center}
\end{figure}


\label{fig:star12.ps}


\label{fig:Period}



\label{fig:Period}



\label{fig:2013.ps}



\end{document}